\magnification=1200
\hsize=125mm
\vsize=185mm
\parindent=8mm
\frenchspacing

\font\Bbb=msbm10
\font\toto=cmbx10 scaled 1400
\font\sect=cmbx10 scaled 1200

\def\ep{\varepsilon}

\def\R{\hbox{\Bbb R}}
\def\S{\hbox{\Bbb S}}
\def\N{\hbox{\Bbb N}}
\def\pa{\partial}
\def\b{\backslash}
\def\A{{\bf A}}

\noindent{\toto On inverse scattering in electromagnetic field
in classical relativistic mechanics at high energies}
\vskip 7mm

\noindent{ Alexandre Jollivet}
\vskip 1cm

\noindent{\bf Abstract.}  We consider the multidimensional Newton-Einstein
equation in static electromagnetic field 
$$
\eqalign{
\dot p = F(x,\dot x),\ F(x,\dot x)=-\nabla V(x)+{1\over c}B(x)\dot x,\cr
p={\dot x \over \sqrt{1-{|\dot x|^2 \over c^2}}},\ \dot p={dp\over dt},\ \dot x={dx\over dt},\ x\in C^1(\R,\R^d),}
\eqno{(*)}
$$
where $V \in C^2(\R^d,\R),$ $B(x)$ is the $d\times d$ real antisymmetric matrix with elements
$B_{i,k}(x)={\pa\over \pa x_i}\A_k(x)-
{\pa\over \pa x_k}\A_i(x)$, and $|\pa^j_x\A_i(x)|+|\pa^j_x V(x)| \le \beta_{|j|}(1+|x|)^{-(\alpha+|j|)}$
for $x\in \R^d,$ $|j| \le 2,$ $i=1..d$ and some $\alpha > 1$.
We give estimates and asymptotics for scattering solutions and scattering data for the equation $(*)$ for the case of small angle
scattering. We show that at high energies the velocity valued component of the scattering operator uniquely determines the X-ray transforms
$P\nabla V$
and $PB_{i,k}$ for $i,k=1..d,$ $i\neq k.$ Applying results on inversion of the X-ray transform $P$ we obtain that for
$d\ge 2$ the velocity valued 
component of the scattering operator at high energies uniquely determines $(V,B)$. In addition we show that
our high energy asymptotics found for the configuration valued component of the scattering operator doesn't determine uniquely $V$ when $d\ge 2$ and 
$B$ when $d=2$ but that it uniquely determines  $B$ when $d\ge 3.$ 

\vskip 1cm

\noindent{\sect 1 Introduction}
\vskip 4mm

\noindent {1.1 \it The Newton-Einstein equation.} 

\noindent Consider the multidimensional 
Newton-Einstein equation in static electromagnetic field 
$$
\eqalign{\dot p = F(x,\dot x),\ F(x,\dot x)=-\nabla V(x)+{1\over c}B(x)\dot x,\cr
p={\dot x \over \sqrt{1-{|\dot x|^2 \over c^2}}},\ \dot p={dp\over dt},\ \dot x={dx\over dt},\ x\in C^1(\R,\R^d),}
\eqno (1.1)
$$
where $V \in C^2(\R^d,\R), B(x)$ is the $d\times d$ real antisymmetric matrix with elements $B_{i,k}(x)={\pa\over \pa x_i}\A_k(x)-{\pa\over
\pa x_k}\A_i(x),$
 $\A=(\A_1,\ldots,\A_d) \in C^2(\R^d,\R^d)$ and
$$|\pa^j_x\A_i(x)|+|\pa^j_x V(x)| \le \beta_{|j|}(1+|x|)^{-(\alpha+|j|)}\eqno (1.2)$$
for $x\in \R^d,$ $|j| \le 2,$ $i=1..d$ and some $\alpha > 1$ (here j is the multiindex $j\in (\N \cup \{0\})^d, |j|= \sum_{n=1}^d j_n$ and $\beta_{|j|}$ 
are positive real constants and $B(x)\dot x=\left(\sum_{l=1}^dB_{1,l}(x)\dot x_l,\ldots,\right.$
$\left.\sum_{l=1}^dB_{d,l}(x)\dot x_l\right)$). The equation (1.1) is an equation for $x=x(t)$ and is the equation of motion in $\R^d$ of a relativistic particle of mass
$m=1$ and charge $e=1$ in an external electromagnetic field described by  $V$ and $\A$
(see [E] and, for example, Section 17 of [LL2]). In this equation $x$ is the position of the particle, $p$ is its impulse, $F$ is the force acting on the particle, $t$ is
the time and $c$ is the speed of light.

For the equation (1.1) the energy
$$
E=c^2\sqrt{1+{|p(t)|^2 \over c^2}}+V(x(t))
$$
is an integral of motion. Note that the energy $E$ does not depend on $\A$ because the magnetic force $(1/c)B(x)\dot x$ is orthogonal to the velocity
$\dot x$ of the particle.

\vskip 4mm
  
\noindent {1.2 \it Yajima's results.}   

\noindent Yajima [Y] studied in dimension 3 (without loss of generality for the case of dimension $d\ge 2)$ the direct scattering of relativistic particle in 
an external electromagnetic field
described by four vector $(V(x),{\bf A}(x))$ where the scalar potential $V$ and the vector potential $\A$ are both rapidly decreasing.
We recall results of Yajima [Y] in our case. We denote by $B_c$ the euclidean open ball whose radius is c and whose centre is 0. 
 
\indent Under the conditions (1.2), the following is valid (see [Y]): for any 
$(v_-,x_-)\in B_c\times\R^{d},\ v_-\neq 0,$
the equation (1.1)  has a unique solution $x\in C^2(\R,\R^d)$ such that
$$
{x(t)=v_-t+x_-+y_-(t),}\eqno (1.3)
$$
where $\dot y_-(t)\to 0,\ y_-(t)\to 0,\ {\rm as}\ t\to -\infty;$  in addition for almost any 
$(v_-,x_-)\in B_c\times \R^{d},\ v_-\neq 0,$
$$
{x(t)=v_+t+x_++y_+(t),}\eqno (1.4)
$$
 where $v_+\neq 0,\ |v_+|<c,\ v_+=a(v_-,x_-),\ x_+=b(v_-,x_-),\ \dot y_+(t)\to 0,\ \ y_+(t)\to 0,{\rm\ as\ }t \to +\infty$.

The map $S: B_c\times\R^d \to B_c\times\R^d $ given by the formulas
$$
{v_+=a(v_-,x_-),\ x_+=b(v_-,x_-)}\eqno(1.5)
$$
is called the scattering map for the equation (1.1). The functions $a(v_-,x_-),$ 

\noindent $b(v_-,x_-)$ are called the scattering data for the equation (1.1).

By ${\cal D}(S)$ we denote the domain of definition of $S$; by ${\cal R}(S)$ we denote the range of $S$ (by definition,
if $(v_-,x_-)\in {\cal D}(S)$, then $v_-\neq 0$ and $a(v_-,x_-)\neq 0$). 

 Under the conditions (1.2), the map $S$  has the following simple properties (see [Y]):
${\cal D}(S)$  is an open subset of $B_c \times \R^d$ and ${\rm Mes}((B_c \times \R^d) \b {\cal D}(S))=0$ for the Lebesgue
measure on $B_c \times \R^d$  induced by the Lebesgue measure on $\R^d\times\R^d$;
the map $S:{\cal D}(S)\to {\cal R}(S)$ is continuous and preserves the element of volume,
$a(v_-,x_-)^2=v_-^2$.
\vskip 4mm

\noindent {1.3 \it A representation of the scattering data.}

\noindent If $V(x)\equiv 0$ and $B(x)\equiv 0,$ then $a(v_-,x_-)=v_-,\ b(v_-,x_-)=x_-,\ (v_-,x_-)\in B_c \times \R^d,\ v_-\neq 0$. Therefore for $a(v_-,x_-),\
b(v_-,x_-)$ we will use the following representation
$$
{\eqalign{
a(v_-,x_-)=&v_-+a_{sc}(v_-,x_-)\cr
b(v_-,x_-)=&x_-+b_{sc}(v_-,x_-)\crcr
}
\hskip 1cm (v_-,x_-)\in {\cal D}(S).}\eqno (1.6)
$$

We will use the fact that, under the conditions (1.2), the map $S$ is uniquely determined by its restriction to ${\cal M}(S)={\cal
D}(S)\cap {\cal M},$ where
$$
{\cal M}=\{(v_-,x_-)\in B_c \times \R^d|v_-\neq 0, v_-x_-=0\}.
$$
This observation is completely similar to the related observation of [No1], [Jo] and is based on the fact that if $x(t)$ satisfies (1.1), then $x(t+t_0)$
also satisfies (1.1) for any $t_0\in\R$.
\vskip 4mm

\noindent {1.4 \it X-ray transform.}

\noindent Consider 
$$
T\S^{d-1}=\{(\theta,x)|\theta \in \S^{d-1},x\in \R^d,\theta x=0\},
$$
where $\S^{d-1}$ is the unit sphere in $\R^d$.

Consider the X-ray transform $P$ which maps each function $f$ with the proper-
ties 
$$
f\in C(\R^d,\R^m),\ |f(x)|=O(|x|^{-\beta}),\ {\rm as\ }|x|\to \infty,{\rm\ for\ some\ }\beta>1
$$
into a function $Pf\in C(T\S^{d-1},\R^m),$ where $Pf$ is defined by
$$
Pf(\theta,x)=\int_{-\infty}^{+\infty}f(t\theta+x)dt,\ (\theta,x)\in T\S^{d-1}.
$$
Concerning the theory of the X-ray transform, the reader is referred to [GGG], [Na], [No1].
\vskip 4mm 

\noindent {1.5 \it Main results of the work.}

\noindent The main results of the present work consist in the small angle scattering estimates for the scattering data $a_{sc}$ and $b_{sc}$ (and scattering
solutions) for the equation (1.1) and in application of these asymptotics and estimates to inverse scattering for the equation (1.1) at high energies.
Our main results include, in particular, Theorem 1.1, Propositions 1.1, 1.2, formulated below in this subsection and Theorems 3.1, 3.2 given in Section
3.
\vskip 2mm
\noindent{\bf Theorem 1.1.}
{\it Let the conditions} (1.2) {\it be valid and}\ $(\theta,x)\in T\S^{d-1}.$
 {\it Let }\ $r$ {\it be a positive constant such that}\ $0<r\le 1,$ $r<c/\sqrt{2}.$   
{\it Then}
$$
\lim\limits_{s\to c\atop s<c}{s\over \sqrt{1-{s^2\over c^2}}}a_{sc}(s\theta,x)
=\int_{-\infty}^{+\infty}F(\tau\theta+x,c\theta)d\tau
,\eqno(1.7{\rm a})
$$
{\it and, in addition,}
$$
\left|\int_{-\infty}^{+\infty}F(\tau\theta+x,s\theta)d\tau-{s\over \sqrt{1-{s^2\over c^2}}}a_{sc}(s\theta,x)\right|
\le {C_1\over \sqrt{1+{s^2\over4(c^2-s^2)}}}\eqno(1.7{\rm b})
$$
{\it for\ }$s_1<s<c,$ {\it where $C_1=C_1(c,d,\beta_0,\beta_1,\beta_2,\alpha,|x|,r)$ and $s_1=s_1(c,d,\beta_1,\beta_2,\alpha,$ $|x|,r)$ are defined in Section}
4 ({\it in subsection} 4.3);
$$
\eqalignno{
\lim\limits_{s\to c\atop s<c}{s^2\over \sqrt{1-{s^2\over c^2}}}b_{sc}(s\theta,x)
=&\int_{-\infty}^0\int_{-\infty}^{\tau}F(\sigma \theta+x,c\theta)d\sigma d\tau
&(1.8{\rm a})\cr
&-\int^{+\infty}_0\int^{+\infty}_{\tau}F(\sigma \theta+x,c\theta)d\sigma d\tau+PV(\theta,x)\theta
,
}
$$
{\it and, in addition,}
$$
\eqalignno{
\left|{b_{sc}(s\theta,x)\over \sqrt{1-{s^2\over c^2}}}\right.-
{1\over c^2} PV(\theta,x)\theta+{1\over s^2}\int_0^{+\infty}\!\!\!\int_{\tau}^{+\infty}F(\sigma\theta+x,s\theta)d\sigma d\tau
\cr
-{1\over s^2}\left.\int^0_{-\infty}\!\!\int^{\tau}_{-\infty}F(\sigma\theta+x,s\theta)d\sigma d\tau\right|
\le  C_2\sqrt{1-{s^2\over c^2}}&&(1.8{\rm b})\cr
}
$$
{\it for\ }$\ s_2<s<c,$ {\it and where $C_2=C_2(c,d,\beta_0,\beta_1,\beta_2,\alpha,|x|,r)$ and $s_2=s_2(c,d,\beta_1,\beta_2,$ $\alpha,|x|,r)$ 
are defined in Section} 4 ({\it in subsection} 4.3).

\vskip 2mm

Theorem 1.1 follows from Theorem 3.1 and Theorem 3.2 given in Section 3.

Consider the vector-functions $w_1(V,\A,\theta,x)$ and $w_2(V,\A,\theta,x),$ 
$(\theta,x)\in$  

\noindent $T\S^{d-1},$ arising in the right-hand sides of
(1.7a) and (1.8a):
$$
\eqalignno{
w_1(V,\A,\theta,x)=&\int_{-\infty}^{+\infty}F(\tau\theta+x,c\theta)d\tau&(1.9{\rm a})\cr
w_2(V,\A,\theta,x)=&\int_{-\infty}^0\int_{-\infty}^{\tau}F(\sigma \theta+x,c\theta)d\sigma d\tau&(1.9{\rm b})\cr
&-\int^{+\infty}_0\int^{+\infty}_{\tau}F(\sigma \theta+x,c\theta)d\sigma d\tau+PV(\theta,x)\theta.\cr
}
$$
\vskip 2mm
\noindent{\bf Remark 1.1.} Using, in particular, that $B$ is antisymmetric one can see that 
the vectors $w_1(V,\A,\theta,x),\ w_2(V,\A,\theta,x)$ are orthogonal to $\theta$
for any $(\theta,x)\in T\S^{d-1}$ and any potential $(V,\A)$ satisfying (1.2).
\vskip 2mm

Let $(V,\A)$ satisfy the conditions (1.2). Define $\tilde{w_1}(V,\A):\R^d\b\{0\}\times \R^d\to \R^d$ by
$$
\eqalign{
\tilde{w_1}(V,\A)(y,x)=&-|y|\int_{-\infty}^{+\infty}\nabla V(sy+x)ds+\int_{-\infty}^{+\infty}B(s y+x)yds\cr
=&|y|w_1(V,\A,{y\over|y|},x-{xy\over |y|^2}y),
}\eqno(1.10)
$$
for $y\in \R^d\b\{0\},\ x\in \R^d.$
Under the conditions (1.2), $\tilde{w_1}(V,\A)=(\tilde{w_1}(V,\A)_1,..,$ 
$\tilde{w_1}(V,\A)_d)$ $\in C^1(\R^d\b\{0\}\times\R^d,\R^d).$ 

Consider the $d$-dimensional smooth manifolds
$$
{\cal V}_{i,k}=\{(\theta,x)\in T\S^{d-1}|\theta_j=0,j=1\ldots d,j\neq i,j\neq k\},\eqno(1.11)
$$
for $i,k=1..d,\ i\neq k.$
\vskip 2mm

\noindent{\bf Proposition 1.1.}
{\it Let $(V,\A)\in C^2(\R^d,\R)\times C^2(\R^d,\R^d)$ satisfy} (1.2). {\it Then

\noindent $w_1(V,\A,\theta,x)$ given for all $(\theta,x)\in T\S^{d-1}$ uniquely determines  $(V,B)$ and the following formulas are valid:
$$
\eqalignno{
P(\nabla V)(\theta,x)=&-{1\over 2}(w_1(V,\A,\theta,x)+w_1(V,\A,-\theta,x)),&(1.12{\rm a})\cr
P(B_{i,k})(\theta,x)=&{1\over 2}\left[{\pa\over\pa y_k}(\tilde{w_1}(V,\A))_i(y,x)+{\pa\over\pa y_k}(\tilde{w_1}(V,\A))_i(-y,x)\right.&(1.12{\rm b})\cr
&\left.-{\pa\over\pa y_i}(\tilde{w_1}(V,\A))_k(y,x)-{\pa\over\pa y_i}(\tilde{w_1}(V,\A))_k(-y,x)\right]_{|y=\theta},\cr}
$$
for $(\theta,x)\in T\S^{d-1},\ i,k=1..d,\ i\neq k;$
$$
\eqalignno{
PB_{i,k}(\theta,x)=&\theta_k{1\over 2}(w_1(V,\A,\theta,x)_i-w_1(V,\A,-\theta,x)_i)&(1.12{\rm c})\cr
&-\theta_i{1\over 2}(w_1(V,\A,\theta,x)_k-w_1(V,\A,-\theta,x)_k)
}
$$
for $(\theta,x)\in {\cal V}_{i,k},\ i,k=1..d,\ i\neq k.$
}
\vskip 2mm

\noindent{\bf Remark 1.2.} Using the formulas (1.12a), (1.12c) and methods of reconstruction of $f$ from $Pf$ (see [GGG], [Na], [No1]), 
$B_{i,k}$ and $V$ can be reconstructed from $w_1(V,\A,\theta,x)$ given for all $(\theta,x)\in {\cal V}_{i,k}$, for $i,k=1..d,\ i\neq k$. 

\vskip 2mm

\noindent{\bf Proposition 1.2.}
{\it Let $(V,\A)\in C^2(\R^d,\R)\times C^2(\R^d,\R^d)$ satisfy} (1.2). {\it Then

\noindent $w_2(V,\A,\theta,x)$ given for all $(\theta,x)\in T\S^{d-1}$ does not determine uniquely $V.$
For $d=2$,
$w_2(V,\A,\theta,x)$ given for all $(\theta,x)\in T\S^{d-1}$ does not determine uniquely $B$}.
{\it For $d\ge3$,
$w_2(V,\A,\theta,x)$ given for all $(\theta,x)\in T\S^{d-1}$ uniquely determines $B$}.
\vskip 2mm

In Section 5 (see Proposition 5.3) we give formulas ((5.17a) and (5.17b)) which show that for $d=3,$ the Fourier transform of the first derivatives of $B$ can be reconstructed from $w_2(V,\A,\theta,x)$ given for all
$(\theta,x)\in T\S^{d-1}$ 
and  we give a formula ((5.17c)) which shows that for $d\ge4$ the X-ray transform of $B$ can be reconstructed from $w_2(V,\A,\theta,x)$ given for all
$(\theta,x)\in T\S^{d-1}$.

Proposition 1.1 and Proposition 1.2 are proved in Section 5.   

From (1.7a) and (1.12) and inversion formulas for the X-ray transform $P$ for $d\ge 2$ (see [R], [GGG], [Na], [No1]) it follows that $a_{sc}$ determines 
uniquely $\nabla V$ and $B$ at high energies. Moreover for $d\ge 2$ methods of reconstruction of $f$ from $Pf$ (see [R], [GGG], [Na], [No1])
permit to reconstruct $\nabla V$ and $B$ from the velocity valued component $a$ of the scattering map at high energies. The formula 
(1.8a) and Proposition 1.2 show that the first term of the asymptotics of $b_{sc}$ doesn't determine
uniquely the potential $V$  when $d\ge 2$ and $B$ when $d=2$ but that it uniquely determines $B$ when $d\ge 3.$ Note that F. Nicoleau paid our attention 
to the fact that, in addition of Proposition 1.2, the vector function $w_2(V,\theta,x), (\theta,x)\in
T\S^{d-1},$ uniquely determines $V$ modulo spherical symmetric potentials when $d\ge 2$, and that $w_2(V,\theta,x), (\theta,x)\in
T\S^{d-1},$ uniquely determines $B$ modulo spherical symmetric magnetic fields when $d=2$. 
 
\vskip 2mm

\noindent{\bf Remark 1.3.} 
The condition (1.2) in all results and estimates which appear in Introduction and in Sections 2, 3, 4 can be weakened to condition (4.11) given at the end of
Section 4. 
\vskip 4mm
\noindent {1.6 \it Historical remarks.}

\noindent Note that inverse scattering for the classical multidimensional Newton equation was first studied by Novikov [No1] without magnetic field (the
existence and uniqueness of the scattering states, asymptotic completness and scattering map for the classical Newton equation were 
studied by Simon [S]). Novikov proved two formulas which link scattering data 
at high energies to the X-ray transform of $-\nabla V$ and $V$. Following Novikov's framework [No1], the author generalized these two formulas to the relativistic 
case without magnetic field in [Jo]. We shall follow the same way to obtain Theorem 1.1 of the present work.
Note also that for the classical multidimensional Newton equation in a bounded open strictly convex domain an inverse boundary value
problem at high energies was first studied in [GN]. 

To our knowledge the inverse scattering problem for a particle in electromagnetic field in classical and classical
relativistic mechanics was not considered in the literature for the case of nonzero magnetic field $B$ before the present article (concerning results given in the literature
on this problem for $B\equiv 0$ see [No1], [Jo] and references therein). However, in quantim mechanics the inverse scattering problem for a particle in electromagnetic
field with $B\not\equiv 0$ was considered, in particular, in [HN], [ER1], [I], [Ju], [ER2], [Ni], [A], [Ha] (concerning results given in the literature on this problem for
$B\equiv 0$ see, in addition, [F], [EW], [No2] and references given in [No2]).
\vskip 4mm
\noindent {1.7 \it Structure of the paper.}

\noindent Further, our paper is organized as follows. In Section 2 we transform the differential equation (1.1) with initial conditions (1.3) into a system of integral
equations which takes the form $(y_-,\dot y_-)=A_{v_-,x_-}(y_-,\dot y_-)$. Then we study $A_{v_-,x_-}$ on a suitable space and we give estimates and contraction estimates
about $A_{v_-,x_-}$ (Lemmas 2.1, 2.2, 2.3).   
In Section 3 we give estimates and asymptotics for the deflection $y_-(t)$ from (1.3) and for  
scattering data $a_{sc}(v_-,x_-),$ $b_{sc}(v_-,x_-)$ from (1.6) (Theorem 3.1 and Theorem 3.2). From these estimates and asymptotics the 
two formulas (1.7a) and (1.8a) will follow when the parameters $c,$ $\beta_m,$ $\alpha,$ $d,$ ${\hat p}_-,$ $x_-$ are fixed and $|v_-|$ increases (where 
$\beta_{|j|},$ $\alpha,$ $d$ are constants from (1.2),
$\beta_m=\max(\beta_0,\beta_1,\beta_2);$ ${\hat p}_-=v_-/|v_-|).$ In these cases $\sup|\theta(t)|$ decreases, where $\theta(t)$ denotes the
angle between the vectors ${\dot x}(t)=v_-+{\dot y}_-(t)$ and $v_-,$ and we deal with small angle scattering. 
Note that, under the conditions of Theorem 3.1, without additional assumptions, there is the estimate $\sup|\theta(t)|<
{1\over4}\pi$ and we deal with rather small angle scattering (concerning the term ``small angle scattering" see [No1] and Section 20 of
[LL1]). Theorem 1.1 follows from Theorem 3.1 and Theorem 3.2. In Section 4 we sketch the proof of Lemmas 2.1, 2.2, 2.3 and Theorem 3.2. 
Section 5 is devoted to Proofs of Proposition 1.1 and Proposition 1.2.

\vskip 4mm

\noindent{\bf Acknowledgement.} This work was fulfilled in the framework of Ph. D. thesis researchs under the direction of R.G. Novikov.

\vskip 1cm

\noindent{\sect 2 A contraction map}
\vskip 4mm
\noindent Let us transform the differential equation (1.1) in a system of integral equations. 
Consider the function $g: \R^d \to B_c$ defined by
$$g(x)={x\over \sqrt{1+{|x|^2\over c^2}}}$$
where $x\in \R^d.$
One can see that $g$ has, in particular, the following simple properties:
$${|g(x)-g(y)|\le \sqrt{d} |x-y|,\ {\rm for\ } x,y\in\R^d,}\eqno(2.1)$$
$g$ is an infinitely smooth diffeomorphism between $\R^d$ and $B_c,$ its inverse is given by
$$\gamma(x)={x\over \sqrt{1-{|x|^2\over c^2}}}, 
x\in B_c.$$ 

Now, if $x$ satisfies the differential equation (1.1) and the initial conditions (1.3), then $x$ satisfies the system of integral
equations 
$$
\eqalignno{
x(t)=&v_-t+x_-+\int\limits_{-\infty}^t\!\left[g\left(\gamma(v_-)+\int\limits_{-\infty}^\tau\! F(x(s),\dot x(s)))ds\right)-v_-\right]d\tau,
&(2.2{\rm a})\cr
\dot x(t)=&g(\gamma(v_-)+\int_{-\infty}^t\! F(x(s),\dot x(s))ds),&(2.2{\rm b})\cr
}
$$
where $F(x,\dot x)=-\nabla V(x)+{1\over c}B(x)\dot x,\ v_-\in B_c\b\{0\}.$

For $y_-(t)$ of (1.3) this system takes the form
$$
(y_-(t),u_-(t))=A_{v_-,x_-}(y_-,u_-)(t),
\eqno(2.3)
$$
where $u_-(t)=\dot y_-(t)$ and
$$
\eqalign{
A_{v_-,x_-}(f,h)(t)=&(A_{v_-,x_-}^1(f,h)(t),A_{v_-,x_-}^2(f,h)(t))\cr
A_{v_-,x_-}^1(f,h)(t)=&\int\limits_{-\infty}^t\!\!\left[g(\gamma(v_-)+\int\limits_{-\infty}^\tau\!\! F(v_-s+x_-+f(s),v_-+h(s))ds)-v_-\right]
d\tau,\cr
A_{v_-,x_-}^2(f,h)(t)=&g(\gamma(v_-)+\int\limits_{-\infty}^t\!\! F(v_-s+x_-+f(s),v_-+h(s))ds)-v_-,
}
$$
for $v_-\in B_c\b\{0\}$.

From (2.3), (1.2) , (2.1) (applied on ``$x$''$=\gamma (v_-)+\int_{-\infty}^{\tau}\!\!F(v_-s+x_-+y_-(s),v_-$ $+\dot y_-(s))ds$ and ``$y$''$=\gamma(v_-)$) and 
$y_-(t)\in C^1(\R,\R^d),\ |y_-(t)|+|\dot y_-(t)|\to 0,$ as $t\to -\infty$, it follows, in particular, that
$$
\eqalign{
(y_-(t),\dot y_-(t))\in& C(\R,\R^d)\times C(\R,\R^d)\cr
{\rm and\ } |\dot y_-(t)|=O(|t|^{-\alpha}),&\ |y_-(t)|=O(|t|^{-\alpha+1}),\ {\rm as\
}t \to -\infty,}
\eqno(2.4)
$$
where $v_-\in B_c\b\{ 0\}$ and $x_-$ are fixed.

Consider the complete metric space
$$
\eqalign{
M_{T,r}=&\{(f,h)\in C(]-\infty,T],\R^d)\times C(]-\infty,T],\R^d)|\ \|(f,h)\|_T\le r\},\cr
{\rm where\ }\|(f,h)\|_T=&\max\left(\sup\limits_{t\in ]-\infty,T]}|h(t)|,\sup\limits_{t\in ]-\infty,T]}|f(t)-t h(t)|\right) \crcr
}\eqno(2.5)
$$
(where for $T=+\infty$ we understand $]-\infty, T]$ as $]-\infty,+\infty[$). From (2.4)
it follows that, at fixed $T<+\infty$,
$$
(y_-(t),\dot y_-(t))\in M_{T,r}  {\rm \ for\ some\ }r\ {\rm depending\ on}\ y_-(t)\ {\rm and}\ T.\eqno(2.6)
$$

Let $z_1(c,d,\beta_1,\alpha,r_x,r)$ be defined as the root of the following equation 
$$
{z_1\over \sqrt{1-{z_1^2\over c^2}}}-{2^{\alpha+5}\beta_1d(2+r/c)\over \alpha({z_1/\sqrt{2}}-r)(r_x/\sqrt{2}+1)^\alpha}=0
,\ z_1\in ]\sqrt{2}r,c[,\eqno(2.7)
$$
where $r_x$ and $r$ are some nonnegative numbers such that $0< r\le 1,\ r<c/\sqrt{2}.$

\vskip 2mm

\noindent{\bf Lemma 2.1.} {\it Under the conditions} (1.2), {\it the following is valid: if} $(f,h)\in M_{T,r},$ $0< r\le 1,$ 
$r<c/\sqrt{2},$
$x_-\in \R^d,$
$v_-\in B_c,$
 $|v_-|
\ge z_1(c,d,\beta_1,\alpha,|x_-|,r),$  
$v_-x_-=0,$ {\it then}
$$
\eqalignno{
\| A_{v_-,x_-}(f,h)\|_T\le& \rho_T(c,d,\beta_1,\alpha,|v_-|,|x_-|,r)&(2.8{\rm a})\cr
=&{1\over \sqrt{1+|v_-|^2/(4(c^2-|v_-|^2))}}\cr
&\times{2^{\alpha+2}d\sqrt{d}\beta_1(2+r/c)(|v_-|/\sqrt{2}+1-r)\over 
 (\alpha-1)(|v_-|/\sqrt{2}-r)^2(1+|x_-|/\sqrt{2}-(|v_-|/\sqrt{2}-r)T)^{\alpha-1}}
\cr
}
$$
{\it for} $T\le 0$,
$$
\eqalignno{
\| A_{v_-,x_-}(f,h)\|_T\le&\rho(c,d,\beta_1,\alpha,|v_-|,|x_-|,r)&(2.8{\rm b})\cr
=&{1\over \sqrt{1+|v_-|^2/(4(c^2-|v_-|^2))}}\cr
&\times{2^{\alpha+3}d\sqrt{d}\beta_1(2+r/c)(|v_-|/\sqrt{2}+1-r)\over 
 (\alpha-1)(|v_-|/\sqrt{2}-r)^2(1+|x_-|/\sqrt{2})^{\alpha-1}}
\cr
}
$$
{\it for} $T\le +\infty$;
{\it if} $\ (f_1,h_1),\ (f_2,h_2)\in M_{T,r},\ 0< r\le 1,\ r<c/\sqrt{2},\ |v_-|<c,$ $v_-x_-=0,$
$|v_-|\ge z_1(c,d,\beta_1,\alpha,|x_-|,r),$ 
{\it then}
$$
\eqalignno{
\| A_{v_-,x_-}(f_2,h_2)&-A_{v_-,x_-}(f_1,h_1)\|_T&(2.9{\rm a})\cr
&\le \lambda_T(c,d,\tilde{\beta},\alpha,|v_-|,|x_-|,r)\|(f_2-f_1,h_2-h_1)\|_T,
}
$$
$$
\eqalign{\lambda_T(c,d,\tilde{\beta},\alpha,|v_-|,|x_-|,r)=&
{1\over \sqrt{1+|v_-|^2/(4(c^2-|v_-|^2))}}\cr
&\times{2^{\alpha+4}d^2\tilde{\beta}(1+{1\over c})({|v_-|\over \sqrt{2}}+1-r)^2\over
(\alpha-1) ({|v_-|\over \sqrt{2}}-r)^3(1+{|x_-|\over \sqrt{2}}-({|v_-|\over \sqrt{2}}-r)T)^{\alpha-1}}\cr
}
$$
{\it for} $T\le 0$,
$$
\| A_{v_-,x_-}(f_2,h_2)-A_{v_-,x_-}(f_1,h_1)\|_T\le \lambda(c,d,\tilde{\beta},\alpha,|v_-|,|x_-|,r)\|(f_2-f_1,h_2-h_1)\|_T,
\eqno(2.9{\rm b})
$$
$$
\eqalign{
\lambda(c,d,\tilde{\beta},\alpha,|v_-|,|x_-|,r)=&{1\over \sqrt{1+|v_-|^2/(4(c^2-|v_-|^2))}}\cr
&\times{2^{2\alpha+9}3d^3\tilde{\beta}(1+\tilde{\beta})(1+1/c)^3(|v_-|/\sqrt{2}+1-r)^3\over
(\alpha-1)(|v_-|/\sqrt{2}-r)^4(1+|x_-|/\sqrt{2})^{\alpha-1}}\cr
}
$$
{\it for} $T\le +\infty,$ {\it where} $\tilde{\beta}=\max(\beta_1,\beta_2).$
\vskip 4mm
Note that 
$$
\max\left({\rho_T(c,d,\beta_1,\alpha,|v_-|,|x_-|,r)\over r}, \lambda_T(c,d,\tilde{\beta},\alpha,|v_-|,|x_-|,r) \right)\ \hskip 3cm
$$
\vskip -6mm
$$
\eqalignno{
\le &\mu_T(c,d,\tilde{\beta},\alpha,|v_-|,|x_-|,r)&(2.10{\rm a})\cr
 =&{1\over \sqrt{1+|v_-|^2/(4(c^2-|v_-|^2))}}\cr
&\times{2^{\alpha+4}d^2\tilde{\beta}(1+1/c)(|v_-|/\sqrt{2}+1-r)^2\over
r(\alpha-1) (|v_-|/\sqrt{2}-r)^3(1+|x_-|/\sqrt{2}-(|v_-|/\sqrt{2}-r)T)^{\alpha-1}}\cr
}
$$
for $T\le 0$,
$$
\max\left({\rho(c,d,\beta_1,\alpha,|v_-|,|x_-|,r)\over r}, \lambda(c,d,\tilde{\beta},\alpha,|v_-|,|x_-|,r)\right)\ \hskip 3cm
$$
\vskip -6mm
$$
\eqalignno{\le &\mu(c,d,\tilde{\beta},\alpha,|v_-|,|x_-|,r)&(2.10{\rm b})\cr
=&{1\over \sqrt{1+|v_-|^2/(4(c^2-|v_-|^2))}}\cr
&\times{2^{2\alpha+9}3d^3\tilde{\beta}(1+\tilde{\beta})(1+1/c)^3(|v_-|/\sqrt{2}+1-r)^3\over
r(\alpha-1)(|v_-|/\sqrt{2}-r)^4(1+|x_-|/\sqrt{2})^{\alpha-1}}
}
$$
for $T\le +\infty$, where $\tilde{\beta}=\max(\beta_1,\beta_2),\ 0< r\le 1,\ r<c/\sqrt{2},\ |v_-|<c,$ $|v_-|
\ge z_1,$
$\ v_-x_-=0.$

From Lemma 2.1 and the estimates (2.10) we obtain the following result.
\vskip 2mm
\noindent{\bf Corollary 2.1.}  {\it Under the conditions} (1.2),$\ 0< r\le 1,$ $r<c/\sqrt{2},$ $x_-\in \R^d,$
$v_-\in B_c,$
$|v_-|\ge$
$z_1(c,d,\beta_1,\alpha,|x_-|,r),\ v_-x_-=0,$ {\it the following result is valid:}

{\it if} $\mu_T(c,d,\tilde{\beta},\alpha,|v_-|,|x_-|,r)<1$, {\it then} $A_{v_-,x_-}$ {\it is a contraction map in }$M_{T,r}$ 
{\it for} $T\le 0;$ 

{\it if} $\mu(c,d,\tilde{\beta},\alpha,|v_-|,|x_-|,r)<1$, {\it then} $A_{v_-,x_-}$ {\it is a contraction map in }$M_{T,r}$ 
{\it for} $T\le +\infty$.
\vskip 4mm
Taking into account (2.6) and using Lemma 2.1, Corollary 2.1 and the lemma about the contraction maps we will study the solution 
$(y_-(t),u_-(t))$ 
of the equation (2.3) in $M_{T,r}$.

We will use also the following results.

\vskip 2mm

\noindent{\bf Lemma 2.2.} {\it Under the conditions} (1.2), $(f,h)\in  M_{T,r},\ 0< r\le 1,\ r<c/\sqrt{2},$ $x_-\in \R^d,$
$v_-\in B_c,$
$|v_-|\ge z_1(c,d,\beta_1,\alpha,|x_-|,r),\ v_-x_-=0,$ {\it the following is valid:}
$$
\eqalignno{
|A_{v_-,x_-}^2(f,h)(t)|\le&\zeta_-(c,d,\beta_1,\alpha,|v_-|,|x_-|,r,t)\cr
=&{1\over \sqrt{1+|v_-|^2/(4(c^2-|v_-|^2))}}&(2.11)\cr
&\times{d\sqrt{d}\beta_12^{\alpha+2}(2+r/c)\over \alpha 
(|v_-|/\sqrt{2}-r)(1+|x_-|/\sqrt{2}-(|v_-|/\sqrt{2}-r)t)^\alpha},\cr
|A_{v_-,x_-}^1(f,h)(t)|\le&\xi_-(c,d,\beta_1,\alpha,|v_-|,|x_-|,r,t)\cr
=&{1\over \sqrt{1+|v_-|^2/(4(c^2-|v_-|^2))}}&(2.12)\cr
&\times{d\sqrt{d}\beta_12^{\alpha+2}(2+r/c)\over \alpha (\alpha-1)
({|v_-|\over\sqrt{2}}-r)^2(1+{|x_-|\over \sqrt{2}}-({|v_-|\over \sqrt{2}}-r)t)^{\alpha-1}},
\crcr
}
$$
{\it for} $t\le T,\ T\le 0;$ 
$$
A_{v_-,x_-}^1(f,h)(t)=k_{v_-,x_-}(f,h)t+l_{v_-,x_-}(f,h)+H_{v_-,x_-}(f,h)(t),\eqno(2.13)
$$
{\it where}
$$
k_{v_-,x_-}(f,h)=g(\gamma (v_-)+\int_{-\infty}^{+\infty}\!\!F(v_-s+x_-+f(s),v_-+h(s))\,ds)-v_-,\eqno(2.14{\rm a})
$$
$$
\eqalignno{
l_{v_-,x_-}(f,h)=&\int_{-\infty}^0\!\left[g(\gamma (v_-)+\int_{-\infty}^{\tau}\! F(v_-s+x_-+f(s),v_-+h(s))\,ds)-v_-\right]\,d\tau
\cr
                   &+\int^{+\infty}_0\!\left[g(\gamma (v_-)+\int_{-\infty}^{\tau}\! F(v_-s+x_-+f(s),v_-+h(s))\,ds)\right.\cr
                   &\left.-g(\gamma (v_-)+\int_{-\infty}^{+\infty}\! F(v_-s+x_-+f(s),v_-+h(s))\,ds)\right]\,d\tau,&(2.14{\rm b})\cr
}
$$
$$
\eqalignno{
|k_{v_-,x_-}(f,h)|\le&2\zeta_-(c,d,\beta_1,\alpha,|v_-|,|x_-|,r,0),&(2.15{\rm a})\cr
|l_{v_-,x_-}(f,h)|\le&2\xi_-(c,d,\beta_1,\alpha,|v_-|,|x_-|,r,0),&(2.15{\rm b})\cr\cr
}
$$
\vskip -8mm
$$
\eqalignno{
|\dot H_{v_-,x_-}(f,h)(t)|\le& \zeta_+(c,d,\beta_1,\alpha,|v_-|,|x_-|,r,t)&(2.16)\cr
                          =&{1\over \sqrt{1+|v_-|^2/(4(c^2-|v_-|^2))}\alpha (|v_-|/\sqrt{2}-r)}\cr
                      &\times{d\sqrt{d}\beta_12^{\alpha+2}(2+r/c) \over
                     (1+|x_-|/\sqrt{2}+(|v_-|/\sqrt{2}-r)t)^{\alpha}},\cr
|H_{v_-,x_-}(f,h)(t)|\le& \xi_+(c,d,\beta_1,\alpha,|v_-|,|x_-|,r,t)&(2.17)\cr
                     =&{1\over \sqrt{1+|v_-|^2/(4(c^2-|v_-|^2))}\alpha (\alpha-1) (|v_-|/\sqrt{2}-r)^2}\cr
                      &\times{d\sqrt{d}\beta_12^{\alpha+2}(2+r/c) \over
                      (1+|x_-|/\sqrt{2}+(|v_-|/\sqrt{2}-r)t)^{(\alpha-1)}},\crcr
}		      
$$
{\it for} $T=+\infty,\ t\ge 0.$

\vskip 4mm
One can see that Lemma 2.2 gives, in particular, estimates and asymptotics for
$$
A_{v_-,x_-}(f,h)(t)=(A_{v_-,x_-}^1(f,h)(t),A_{v_-,x_-}^2(f,h)(t))\ \ {\rm as}\  t\to\pm\infty.
$$
\vskip 2mm
\noindent{\bf Lemma 2.3.} {\it Let the conditions} (1.2) {\it be valid,} $(y_-(t),u_-(t))\in  M_{T,r}$ {\it be a solution of\ } (2.3), $T=+\infty,
\ 0< r\le 1,\ r<c/\sqrt{2},$ $x_-\in \R^d,$
$v_-\in B_c,$ $|v_-|\ge z_1(c,d,\beta_1,\alpha,|x_-|,r),$ $v_-x_-=0,$ {\it then}
$$
\eqalignno{
|k_{v_-,x_-}(y_-,u_-)-k_{v_-,x_-}(0,0)|\le & \ep_a'(c,d,\beta_1,\tilde{\beta},\alpha,|v_-|,|x_-|,r)\cr
=&{d^2\tilde{\beta}(1+{1\over c})2^{\alpha +5}(|v_-|/\sqrt{2}+1-r)\over
\alpha (|v_-|/\sqrt{2}-r)^2(1+|x_-|/\sqrt{2})^\alpha}&(2.18{\rm a})\cr
& \times {\rho(c,d,\beta_1,\alpha,|v_-|,|x_-|,r) \over \sqrt{1+|v_-|^2/(4(c^2-|v_-|^2))}},
}
$$
$$
\eqalignno{
\left|{k_{v_-,x_-}(y_-,u_-)\over\sqrt{1-{|v_-|^2\over c^2}}}-\int_{-\infty}^{+\infty}\!\!\!\!\!\!\!\!F(x_-+v_-s,v_-)\,ds\right|
\le&\ep_a(c,d,\beta_1,\tilde{\beta},\alpha,|v_-|,|x_-|,r)&(2.18{\rm b})\cr
=&{2^{\alpha +5}d\sqrt{d}\tilde{\beta}(1+{1\over c})(|v_-|/\sqrt{2}+1-r)\over
\alpha (|v_-|/\sqrt{2}-r)^2(1+|x_-|/\sqrt{2})^\alpha}\cr
&\times\rho(c,d,\beta_1,\alpha,|v_-|,|x_-|,r),
}
$$
$$
\eqalignno{
|l_{v_-,x_-}(y_-,u_-)-l_{v_-,x_-}(0,0)|\le & \ep_b(c,d,\beta_1,\tilde{\beta},\alpha,|v_-|,|x_-|,r)&(2.18{\rm c})\cr
=&{2^{2\alpha+9}d^3\tilde{\beta}(1+\tilde{\beta})3(1+1/c)^3(|v_-|/\sqrt{2}+1-r)^2\over
(\alpha-1)(|v_-|/\sqrt{2}-r)^4(1+|x_-|/\sqrt{2})^{\alpha-1}}
\cr
& \times{\rho(c,d,\beta_1,\alpha,|v_-|,|x_-|,r) \over \sqrt{1+|v_-|^2/(4(c^2-|v_-|^2))}},\crcr
}
$$
{\it where} $k_{v_-,x_-}$ {\it and} $l_{v_-,x_-}$ {\it are defined in} (2.14) {\it and} $\rho$ {\it is defined in} (2.8b).
\vskip 2mm
We sketch the proof of Lemmas 2.1, 2.2, 2.3 in Section 4.
\vskip 1cm

\noindent{\sect 3 Small angle scattering}
\vskip 4mm

\noindent Under the conditions (1.2), for any $(v_-,x_-)\in B_c\times \R^d,\ v_-\neq 0,$ the equation (1.1) has a unique solution $x \in
C^2(\R,\R^d)$ with the initial conditions (1.3). Consider the function $ y_-(t)$ from (1.3). This function describes deflection from 
free motion.

Using Corollary 2.1 the lemma about contraction maps, and Lemmas 2.2 and 2.3 we obtain the following result.
\vskip 2mm
\noindent{\bf Theorem 3.1.}
{\it Let the conditions} (1.2) {\it be valid},
$\mu(c,d,\tilde{\beta},\alpha,|v_-|,|x_-|,r)<1$, $\tilde{\beta}=\max(\beta_1,\beta_2),\ 0< r\le 1,\ r<c/\sqrt{2},$ $x_-\in\R^d,$ $v_-\in B_c,$ $|v_-|\ge
z_1(c,d,\beta_1,\alpha,|x_-|,r),$  $v_-x_-=0$, {\it where} $\mu$ {\it is defined by} (2.10b) {\it and} 
$z_1$ {\it is defined by} (2.7). {\it Then the deflection} $y_-(t)$ {\it has the following
properties:}
$$(y_-,\dot y_-)\in M_{T,r},\ T=+\infty;\eqno(3.1)$$
$$
\eqalignno{
|{\dot y}_-(t)|\le&\zeta_-(c,d,\beta_1,\alpha,|v_-|,|x_-|,r,t),&(3.2)\cr
|y_-(t)|\le& \xi_-(c,d,\beta_1,\alpha,|v_-|,|x_-|,r,t)\ \ {\it for}\ \ t\le 0;
&(3.3)\cr
y_-(t)=&a_{sc}(v_-,x_-)t + b_{sc}(v_-,x_-) + h(v_-,x_-,t),&(3.4)\crcr
}
$$
{\it where}
$$
\left|a_{sc}(v_-,x_-)-\left[{\gamma(v_-) +\int_{-\infty}^{+\infty}
F(v_-s + x_-,v_-)ds \over \sqrt{1+{|\gamma(v_-) +\int_{-\infty}^{+\infty}F( v_-s + x_-,v_-)ds|^2\over c^2}}}-v_-\right]\right|
$$
$$
\le \ep_a'(c,d,\beta_1,\tilde{\beta},\alpha,|v_-|,|x_-|,r),\eqno(3.5{\rm a})
$$
$$
\left|{a_{sc}(v_-,x_-)\over \sqrt{1-{|v_-|^2\over c^2}}}-\int_{-\infty}^{+\infty}F(v_-s + x_-,v_-)ds\right|
\le \ep_a(c,d,\beta_1,\tilde{\beta},\alpha,|v_-|,|x_-|,r),\eqno(3.5{\rm b})
$$
$$
|b_{sc}(v_-,x_-)-l_{v_-,x_-}(0,0)|\le \ep_b(c,d,\beta_1,\tilde{\beta},\alpha,|v_-|,|x_-|,r),\eqno(3.5{\rm c})
$$
$$
\eqalignno{
|a_{sc}(v_-,x_-)|\le& 2\zeta_-(c,d,\beta_1,\alpha,|v_-|,|x_-|,r,0),&(3.6{\rm a})\cr
|b_{sc}(v_-,x_-)|\le& 2\xi_-(c,d,\beta_1,\alpha,|v_-|,|x_-|,r,0),& (3.6{\rm b})\cr
|{\dot h}(v_-,x_-,t)|\le& \zeta_+(c,d,\beta_1,\alpha,|v_-|,|x_-|,r,t),&(3.7)\cr
|h(v_-,x_-,t)|\le& \xi_+(c,d,\beta_1,\alpha,|v_-|,|x_-|,r,t),&(3.8)\cr
}
$$
{\it for} $t\ge 0$, {\it where} $l_{v_-,x_-}(0,0)$ {\it (resp.} $\ep_a',$ $\ep_a,$ $\ep_b,$ $\zeta_-,$ $\zeta_+,$ $\xi_-$ {\it and} $\xi_+$)
{\it is defined in} (2.14b) ({\it resp.} (2.18a), (2.18b), (2.18c), (2.11), (2.16), (2.12) {\it and} (2.17)).

\vskip 4mm

Let $z=z(c,d,\tilde{\beta},\alpha,r_x,r)$ and $z_2=z_2(c,d,\beta_1,\alpha,r_x)$ be defined as the roots of the
following equations
$$
\mu(c,d,\tilde{\beta},\alpha,z,r_x,r)=1,\ z\in]\sqrt{2}r,c[,\eqno(3.9)
$$
$$
{z_2\over \sqrt{1-{z_2^2\over c^2}}}-{32\beta_1 d\over \alpha (z_2/\sqrt{2})(1+r_x/\sqrt{2})^{\alpha}}=0,\ z_2\in ]0,c[,
\eqno(3.10)
$$
where $\mu$ is defined by (2.10b), $r_x$ and $r$ are some nonnegative numbers such that $0< r\le 1,\ r<c/\sqrt{2},$  
and where $\tilde{\beta}=\max(\beta_1,\beta_2).$

We use the following observations.

(I) Let $\ 0< r\le 1, r<c/\sqrt{2},\ 0\le \sigma$
$$
{s_1\over \sqrt{1-{s_1^2\over c^2}}}-{2^{\alpha+5}\beta_1d(2+r/c)\over \alpha({s_1/\sqrt{2}}-r)(\sigma/\sqrt{2}+1)^\alpha}
>
{s_2\over \sqrt{1-{s_2^2\over c^2}}}-{2^{\alpha+5}\beta_1d(2+r/c)\over \alpha({s_2/\sqrt{2}}-r)(\sigma/\sqrt{2}+1)^\alpha}
$$
for $\sqrt{2}r <s_2<s_1<c$.

(II) Let $\ 0< r\le 1, r<c/\sqrt{2},\ \sigma\in ]\sqrt{2}r,c[,$
$$
{\sigma\over \sqrt{1-{\sigma^2\over c^2}}}-{2^{\alpha+5}\beta_1d(2+r/c)\over \alpha({\sigma/\sqrt{2}}-r)(s_1/\sqrt{2}+1)^\alpha}
>
{\sigma\over \sqrt{1-{\sigma^2\over c^2}}}-{2^{\alpha+5}\beta_1d(2+r/c)\over \alpha({\sigma/\sqrt{2}}-r)(s_2/\sqrt{2}+1)^\alpha}
$$
for $0\le s_2<s_1$. 

(III) Let $0< r\le 1,\ r<c/\sqrt{2},$ $x$ some real nonnegative number, $\tilde{\beta}=\max(\beta_1,\beta_2)$ and $\sqrt{2}r<s<c$ then
$$
\mu(c,d,\tilde{\beta},\alpha,s,|x|,r)<1\Leftrightarrow s>z(c,d,\tilde{\beta},\alpha,|x|,r).
$$
Observations (I) and (II) imply that  $z_1(c,d,\beta_1,\alpha,s_2,r)>z_1(c,d,\beta_1,\alpha,s_1,r)$ for $\sqrt{2}r<s_2<s_1<c$ when $c,\ \beta_1,\ \alpha,
\  d,\ r$ are fixed. 

Theorem 3.1 gives, in particular, estimates for the scattering process and asymptotics for the velocity valued component of
the scattering map when $c,$ $\beta_1,$ $\beta_2,$ $\alpha,$ $d,$ $\hat v_-,$ $x_-$ are fixed (where $\hat v_-=v_-/|v_-|$) and $|v_-|$ increases or, 
e.g., $c,$ $\beta_1,$ $\beta_2,$ $\alpha,$  $d,$ $v_-,$ $\hat x_-$ are fixed and $|x_-|$ increases. In these cases $\sup_{t\in \R}|\theta(t)|$ 
decreases, where $\theta(t)$ denotes the angle between the vectors $\dot x(t)=v_-+\dot y_-(t)$ and $v_-$, and we deal with small angle 
scattering. Note that already under the conditions of Theorem 3.1, without additional assumptions, there is the estimate 
$\sup_{t\in \R}|\theta(t)|<{1\over 4}\pi$ and we deal with a rather small angle scattering. 
Theorem 3.1 with (3.5c) will give the asymptotics of the configuration valued component $b(v_-,x_-)$ of the scattering map if we can study the 
asymptotics of $l_{v_-,x_-}(0,0)$. This is the subject of Theorem 3.2.

\vskip 2mm

\noindent{\bf Theorem 3.2.} {\it Let} $c,\ d,\ \beta_0,\ \beta_1,\ \alpha,\ |x|$ {\it be fixed. Then there exists  a constant} 
$C_{c,d,\beta_0,\beta_1,\alpha,|x|}$ {\it \ such that\ }
$$
\eqalignno{
\left|{l_{v,x}(0,0)\over \sqrt{1-{|v|^2\over c^2}}}
-{1\over c^2} PV({\hat v},x){\hat v}\right.+{1\over|v|^2}&\int_0^{+\infty}\!\!\!\int_{\tau}^{+\infty}F(\sigma{\hat v}+x,v)d\sigma d\tau
\cr
\left.-{1\over|v|^2}\int^0_{-\infty}\!\!\int^{\tau}_{-\infty}F(\sigma{\hat v}+x,v)d\sigma d\tau\right|\le&C_{c,d,\beta_0,\beta_1,\alpha,|x|}
\sqrt{1-{|v|^2\over c^2}}&(3.11) 
}
$$
{\it for any }$\ v\in B_c,$ $|v|\ge z_2(c,d,\beta_1,\alpha,|x|),$ $vx=0,$ {\it and where }
${\hat v}=v/|v|.$
\vskip 2mm
We sketch the proof of Theorem 3.2 in Section 4.
\vskip 1cm

\noindent{\sect 4 About the proof of Lemmas 2.1, 2.2, 2.3 and Theorems 3.2 and 1.1}
\vskip 4mm
 
\noindent The way we prove Lemmas 2.1, 2.2, 2.3 and Theorem 3.2 of the present work, is actually exactly the same as the way we prove lemmas 2.1, 2.2, 2.3 
and theorem 3.2 of [Jo].
\vskip 4mm
\noindent 4.1 {\it Inequalities for} $F$ {\it and} $g$.

\noindent Before sketching the proof of Lemmas 2.1, 2.2, 2.3 and Theorem 3.2, we shall give some estimates about the growth of $g$ defined by
$$
g(x)={x\over \sqrt{1+{|x|^2\over c^2}}},\ x\in\R^d,
$$
and we shall prove Lemma 4.1 given below.

We remind that $g$ has the following simple properties (see [Jo]):
$$
\eqalignno{
|\nabla g_i(x)|^2\le& {1\over{1+{|x|^2\over c^2}}},&(4.1)\cr
|g(x)-g(y)|\le& \sqrt{d} \sup\limits_{\ep\in [0,1]}{1\over\sqrt{1+{|\ep x+(1-\ep) y|^2\over c^2}}}|x-y|,
&(4.2)\cr
|\nabla g_i(x)-\nabla g_i(y)|\le& {3 \sqrt{d}\over c} \sup\limits_{\ep\in [0,1]}{1\over1+{|\ep x+(1-\ep) y|^2\over c^2}} |x-y|,
&(4.3)\crcr
}
$$ 
{\it for } $x,\ y\in \R^d,\ i=1..d,$ and where $g=(g_1,..,g_d).$
\vskip 2mm

\noindent{\bf Lemma 4.1.}  {\it Under the conditions} (1.2), {\it the following estimates are valid:}
 $$
\eqalignno{
|F(x,y)|\le&2d\beta_1(1+|x|)^{-(\alpha+1)}(1+{1\over c}|y|)
\ {\it for\ }x,y\in\R^d,
&(4.4)\cr
|F(x,y)-F(x',y')|\le &{1\over c}2d\beta_1 \sup_{\ep\in [0,1]}(1+|\ep x+(1-\ep) x'|)^{-(\alpha+1)}|y-y'|&(4.5)\cr
+2d\sqrt{d}\beta_2&\sup_{\ep \in[0,1]}(1+|\ep y+(1-\ep) y'|/c)(1+|\ep x+(1-\ep) x'|)^{-(\alpha+2)}\cr
\times |x-x'|,\cr
}
$$
{\it for } $x,\ y,\ x',\ y'\in \R^d.$

{\it Let} $(f,h),\ (f_1,h_1),\ (f_2,h_2)\in M_{T,r},\ v_-\in B_c\b\{0\},\ v_-x_-=0,\ |v_-|>\sqrt{2}r,$ {\it then}
$$
\eqalignno{
|f(s)|\le&(1+|s|)\|(f,h)\|_T,&(4.6)\cr
|h(s)|\le&\|(f,h)\|_T,&(4.7)
}
$$
for $s\le T;$
$$
2(1+|x_-+v_-s+f(s)|)\ge (1+|x_-|/\sqrt{2}+(|v_-|/\sqrt{2}-r)|s|),{\it \ for\ }s\le T,\eqno(4.8)
$$
$$
\left|\int_{-\infty}^t F(v_-s+x_-+f(s),v_-+h(s))ds\right|\le {\beta_1d2^{\alpha+3}(2+r/c)\over \alpha (|v_-|/\sqrt{2}-r)(|x_-|/\sqrt{2}+1)^\alpha}
,\eqno(4.9)
$$
$$
\eqalignno{
&\left(1+{1\over c^2}\left|\gamma(v_-)+\ep_1 \int_{-\infty}^tF(v_-s+x_-+f_1(s),v_-+h_1(s))ds\right.\right.\cr
&\left.\left.+\ep_2
\int_{\sigma}^\tau F(v_-s+x_-+f_2(s),v_-+h_2(s))ds\right|^2\right)
^{-\beta}\cr
&\le (1+{|v_-|^2\over 4(c^2-|v_-|^2)})^{-\beta},&(4.10)
}
$$
{\it for} $\tau,t\in ]-\infty, T],\ \sigma\in[-\infty,\tau],$ $\beta> 0,$ $-1\le\ep_1\le 1,$ $-1\le\ep_2\le 1,$ $(f_1,h_1),$ $(f_2,h_2)\in M_{T,r}$ {\it and if} 
$|v_-|\ge z_1(c,d,\beta_1,\alpha,|x_-|,r),$ $|v_-|<c,$ {\it where} $\gamma$ {\it is defined by}
$$
\gamma(v)={v\over \sqrt{1-|v|^2/c^2}},
$$
{\it for} $v\in B_c.$
\vskip 2mm
\noindent{\it Proof of Lemma} 4.1.
The estimates (4.4) and (4.5) follows directly from the formula $F(x,y)=-\nabla V(x)+{1\over c}B(x)y$ and 
$B(x)=[{\pa\over \pa x_j}\A_k(x)-{\pa\over \pa x_k}\A_j(x)]_{j,k=1..d}$ and the conditions (1.2). The inequalities (4.6), 
(4.7) and (4.8) follow from the definition of $M_{T,r}.$ Using (4.4),
(4.8), we obtain (4.9) and using (4.9) and  the definition of $z_1(c,d,\beta_1,\alpha,|x_-|,r)$ we obtain (4.10).
\hfill $\sqcap\hskip -2.32mm\sqcup$

\vskip 4mm

\noindent 4.2 {\it Sketch of proofs of Lemmas 2.1, 2.2, 2.3 and Theorem 3.2.}

\noindent One can prove Lemmas 2.1, 2.2, 2.3 of the present work by repeating the proof of lemmas 2.1, 2.2, 2.3 of [Jo] and by making the following replacements.
First the estimates given in lemmas 4.1, 4.3 of [Jo] are replaced by the estimates of Lemma 4.1 of the present work. 
Then, to prove Lemmas 2.1, 2.2, 2.3, we replace ${d\over dt}A_{v_-,x_-}(f),$   $A_{v_-,x_-}(f)$ and $F(v_-s+x_-+f(s)),$ for $f\in
M_{T,r},$ in the proof of lemmas 2.1, 2.2,
2.3 of [Jo] by
$ A_{v_-,x_-}^2(f,h),$ $A_{v_-,x_-}^1(f,h)$ and  $F(v_-s+x_-+f(s), v_-+h(s))$ for $(f,h)\in M_{T,r}.$
\hfill $\sqcap\hskip -2.32mm\sqcup$

One can prove Theorem 3.2 by repeating the proof of theorem 3.2 of [Jo] and by making the following replacements. 
We replace the estimates given in lemmas 4.1, 4.3 of [Jo] by the estimates of Lemma 4.1 of our
present work and we replace $F(\tau\theta +x)$ 
of the proof of theorem 3.2 of [Jo] by $F(\tau \theta+x, s\theta).$
\hfill $\sqcap\hskip -2.32mm\sqcup$

\vskip 4mm
\noindent 4.3 {\it Constants} $C_1,$ $C_2,$ $s_1,$ $s_2$ {\it of Theorem 1.1.}

\noindent As it was mentioned already in Introduction, Theorem 1.1 follows from Theorem 3.1 and Theorem 3.2. In addition, constants $C_1,$ $C_2,$ 
$s_1,$ $s_2$, which appear in Theorem 1.1, are given explicitly by
$$
\eqalign{
s_1=&\max(z(c,d,\tilde{\beta},\alpha,|x|,r),z_1(c,d,\beta_1,\alpha,|x|,r)), \cr
s_2=&\max(z(c,d,\tilde{\beta},\alpha,|x|,r),z_1(c,d,\beta_1,\alpha,|x|,r),z_2(c,d,\beta_1,\alpha,|x|)),\cr
C_1=&{d^3\tilde{\beta}^22^{2\alpha+9}(1+{1\over c})^2c({c\over \sqrt{2}}+1-r)^2\over \alpha (\alpha-1)
({s_1\over \sqrt{2}}-r)^4(1+{|x|\over \sqrt{2}})^{2\alpha-1}},\cr
C_2=&C_{c,d,\beta_0,\beta_1,\alpha,|x|}+
{4d^4\sqrt{d}\tilde{\beta}^2(1+\tilde{\beta})2^{3\alpha+15}(1+1/c)^4({c\over \sqrt{2}}+1-r)^3\over
(\alpha-1)^2({s_2\over \sqrt{2}}-r)^6(1+{|x|\over \sqrt{2}})^{2\alpha-2}},\cr
}
$$
where $C_{c,d,\beta_0,\beta_1,\alpha,|x|}$ is the constant of Theorem 3.2 and $z$, $z_1$, $z_2$ are defined by (3.9), (2.7), (3.10)
and where $\tilde{\beta}=\max(\beta_1,\beta_2)$.

\vskip 4mm
\noindent 4.4 {\it Weakened assumptions.} 

\noindent Let $M_d(\R)$ denote the space of $d\times d$ matrix with real elements.
Let $V \in C^2(\R^d,\R)$ so that:
$$
|\pa^j_x V(x)| \le \beta_{|j|}(1+|x|)^{-(\alpha+|j|)},\eqno(4.11{\rm a})
$$
for $|j| \le 2$ and some $\alpha > 1$ (here j is the multiindex $j\in (\N \cup \{0\})^d, |j|= \sum_{n=1}^d j_n$ and $\beta_{|j|}$ 
are positive real constants).
Let $B\in C^1(\R^d,M_d(\R))$ so that: 
$$
\eqalignno{
B(x)\ {\rm is\ a\ } d\times d\ {\rm antisymmetric\ matrix\ with\ real\ }&{\rm elements\ }B_{m,n}(x),&(4.11{\rm b})\cr
{\pa\over \pa x_i}B_{k,l}(x)+{\pa\over \pa x_l}B_{i,k}(x)+{\pa\over \pa x_k}B_{l,i}(x)&=0,&(4.11{\rm c})
}
$$
for $x\in\R^d,$ for $i,k,l=1..d$; 
$$
|\pa^j_x B_{i,k}(x)|\le  \beta_{|j|+1}(1+|x|)^{-(\alpha+|j|+1)},\eqno(4.11{\rm d})
$$
for $i,k=1..d$ and for $|j| \le 1$. 

Let $\A$ be the transversal gauge of $B$, i.e. 
$$
\A(x)=-\int_0^1sB(sx).x ds.\eqno(4.12)
$$
Under the conditions (4.11b), (4.11c) and (4.11d), $\A$ satisfies
$$
\eqalignno{
|\A(x)|\le& \beta (1+|x|)^{-1},&(4.13{\rm a})\cr
B_{i,k}(x)=&{\pa\over \pa x_i}\A_k(x)-
{\pa\over \pa x_k}\A_i(x).&(4.13{\rm b})
}
$$
for $x\in \R^d,$ $i,k=1..d$ and some positive real constant $\beta.$  

If we replace assumptions (1.2) by assumptions (4.11) given above, then the estimates (4.4) and (4.5) still hold. As a consequence, using also Remark 5.2,
we obtain that assumptions (1.2) in all results and estimates which appear in Introduction and in Sections 2, 3 can be weakened to assumptions (4.11).

\vskip 1cm

\noindent{\sect 5 Proofs of Proposition 1.1 and Proposition 1.2}

\vskip 4mm
\noindent Let $\A \in C^2(\R^d,\R^d)$ and
$$
|\pa^j_x \A_i(x)|\le \beta_{|j|}(1+|x|)^{-(\alpha+|j|)}\eqno(5.1)
$$
for $x\in \R^d,$ $|j| \le 2, \ i=1..d$ and some $\alpha > 1$ (here j is the multiindex $j\in (\N \cup \{0\})^d, |j|= \sum_{n=1}^d j_n$ and 
$\beta_{|j|}$ are positive real constants).
We define the magnetic field $B\in C^1(\R^d,{\cal M}_d(\R))$ by:
$B(x)$ is the $d\times d$ real antisymmetric matrix with elements
$$
B_{i,k}(x)={\pa \over \pa x_i}\A_k(x)-{\pa \over \pa x_k}\A_i(x)
\eqno(5.2)
$$ 
for $x\in \R^d$ (where ${\cal M}_d(\R)$ denotes the space of $d\times d$ real matrix).
For  $\A\in C^2(\R^d,\R^d)$ satisfying (5.1) and $(\theta,x)\in T\S^{d-1}$ we define the vectors $w_3(\A,\theta,x)$ and $w_4(\A,\theta,x)$:
$$
\eqalignno{
w_3(\A,\theta,x)=&\int_{-\infty}^{+\infty}\!\!\!\!\! B(x+\sigma\theta)\theta d\sigma,&(5.3{\rm a})\cr
w_4(\A,\theta,x)=&\int_{-\infty}^0\!\int_{-\infty}^\tau\!\!\!\!\! B(x+\sigma\theta)\theta d\sigma d\tau
-\int^{+\infty}_0\!\!\!\int^{+\infty}_\tau\!\!\!\!\! B(x+\sigma\theta)\theta d\sigma d\tau,&(5.3{\rm b})
}
$$
where $B$ is defined by (5.2).

We also define a function $\tilde{w_3}(\A):\R^d\b\{0\}\times\R^d\to\R^d$ by
$$
\tilde{w_3}(\A)(y,x)=|y|w_3(\A,{y\over|y|},x-{xy\over|y|^2}y),\eqno(5.4)
$$
for $x\in\R^d,y\in \R^d\b\{0\}.$
From (5.1), (5.4) and (5.3a) it follows that
$$
\tilde{w_3}(\A)(y,x)=\int_{-\infty}^{+\infty}B(x+\sigma y)yd\sigma,\eqno(5.5)
$$
for $(x,y)\in\R^d\times\R^d\b\{0\}.$
From (5.1), it follows that 
$
\tilde{w_3}(\A)=((\tilde{w_3}(\A))_1,..,$ $(\tilde{w_3}(\A))_d)
\in$ $ C^1(\R^d\b\{0\}\times \R^d,\R^d).
$

To prove Proposition 1.1 we first prove the following result.

\vskip 2mm

\noindent{\bf Proposition 5.1.} 
{\it Let $\A \in C^2(\R^d,\R^d)$ satisfy} (5.1) {\it . 
Then $w_3(\A,\theta,x)$ given for all $(\theta,x)\in T\S^{d-1}$ determines uniquely the magnetic field $B$ defined by }
 (5.2) {\it and the following formulas are valid:
$$
PB_{i,k}(\theta,x)=\left({\pa\over\pa y_k}(\tilde{w_3}(\A))_i(y,x)-{\pa\over\pa y_i}(\tilde{w_3}(\A))_k(y,x)\right)_{|y=\theta},\eqno(5.6{\rm a})
$$
for $(\theta,x)\in T\S^{d-1},\ i,k=1..d,\ i\neq k;$
$$
\eqalignno{
PB_{i,k}(\theta,x)=&\theta_k w_3(\A,\theta,x)_i-\theta_i w_3(\A,\theta,x)_k&(5.6{\rm b})
}
$$
for $(\theta,x)\in {\cal V}_{i,k},\ i,k=1..d,\ i\neq k$ where ${\cal V}_{i,k}$ is the $d$-dimensional smooth manifold given by} (1.11).
\vskip 2mm

Note that under different conditions on  vector potentials $\A,$ the question of the determination of $B$ from $w_3$ was studied in [Ni], [Ju], [I]. 
However, to our knowledge the formulas (5.6) were not given in the literature. 

\vskip 2mm

\noindent{\it Proof of Proposition} 5.1.
Under the conditions (5.1) and from (5.2) and (5.5) it follows that  
$$
\eqalignno{
{\pa\over\pa y_k}(\tilde{w_3}(\A))_i(y,x)=&\int_{-\infty}^{+\infty}[{\pa\over \pa x_i}\A_k(ty+x)-{\pa\over \pa x_k}\A_i(ty+x)]dt
&(5.7)\cr
&+\sum_{j=1}^d\int_{-\infty}^{+\infty}t[{\pa^2\over \pa{x_k}\pa x_i}\A_j(ty+x)-
{\pa^2\over \pa x_k\pa x_j}\A_i(ty+x)]y_j dt
}
$$
for any $(y,x)\in \R^d\b\{0\}\times\R^d$ and $i,k=1..d$.
Let $i,k=1..d.$ From (5.7) it follows that
$$ 
\eqalignno{
\left({\pa\over\pa y_k}(\tilde{w_3}(\A))_i(y,x)\right.&-\left.{\pa\over\pa y_i}(\tilde{w_3}(\A))_k(y,x)\right)_{|y=\theta}\cr
=&2PB_{i,k}(\theta,x)+\int_{-\infty}^{+\infty}t \sum_{j=1}^d {\pa\over \pa x_j}B_{i,k}(t\theta+x) \theta_j dt,&(5.8)\cr
}
$$
for all $(\theta,x)\in T\S^{d-1},\ \theta=(\theta_1,\ldots,\theta_d)$ and where $ P$ denotes the X-ray transform. 
Integrating by parts the integral of the right-hand side of (5.8), we
obtain the formula (5.6a).

We recall that
$w_3(\A,\theta,x)_i=\sum_{j=1}^d\int_{-\infty}^{+\infty}B_{i,j}(t\theta+x)\theta_j dt,$
for $(\theta,x)\in T\S^{d-1},$ $i,k=1..d,$ $i\neq k$.
Hence $\theta_kPB_{i,k}(\theta,x)=w_3(\A,\theta,x)_i$
for $(\theta,x)\in {\cal V}_{i,k},$ $i,k=1..d,$ $i\neq k$. This last formula implies
(5.6b)  ($\theta_i^2+\theta_k^2=1$ for $(\theta,x)\in {\cal V}_{i,k},$ $\theta=(\theta_1,\ldots,\theta_d)$).

Then using results on inversion of the X-ray transform and using (5.6a) or (5.6b) and using (5.4) we obtain that
$w_3(\A,\theta,x)$ given for all $(\theta,x)\in T\S^{d-1}$ uniquely determines the magnetic field $B.$

Proposition 5.1 is proved.
\hfill $\sqcap\hskip -2.32mm\sqcup$
\vskip 2mm

Now we are ready to prove Proposition 1.1.

Let $(\theta,x)\in T\S^{d-1}.$
We note that
$$
\int_{-\infty}^{+\infty} B(\tau(-\theta)+x)(-\theta) d\tau=-\int_{-\infty}^{+\infty}
 B(\tau\theta+x)\theta d\tau \eqno(5.9{\rm a})
$$
and we remind that
$$
P(\nabla V)(-\theta,x)=P(\nabla V)(\theta,x).\eqno(5.9{\rm b})
$$
Using (1.9a), (5.3a) and (5.9) we obtain the formula (1.12a) and the following formula
$$
w_3(\A,\theta,x)={1\over 2}(w_1(V,\A,\theta,x)-w_1(V,\A,-\theta,x)),\eqno(5.10)
$$
for $(\theta,x)\in T\S^{d-1}.$
From (1.12a) and results on inversion of the X-ray transform, we obtain that 
$w_1(V,\A,\theta,x)$ given for all $(\theta,x)\in T\S^{d-1}$ uniquely determines $\nabla V$ and thus it uniquely determines $V$ (
$(V,\A)$ satisfies (1.2)). From (5.10) and Proposition 5.1 it follows that  
$w_1(V,\A,\theta,x)$ given for all $(\theta,x)\in T\S^{d-1}$ uniquely determines $B$. In addition from (5.10) it follows that
$$
\tilde{w_3}(\A)(y,x)={1\over 2}(\tilde{w_1}(V,\A)(y,x)-\tilde{w_1}(V,\A)(-y,x)),
$$
for $y\in\R^d\b\{0\},\ x\in \R^d.$ Using this last formula and (5.6a) of Proposition 5.1 we obtain (1.12b).
Using (5.10) and (5.6b), we obtain (1.12c).

Proposition 1.1 is proved.
\hfill $\sqcap\hskip -2.32mm\sqcup$
\vskip 2mm

Let $i,k=1..d,\ i\neq k.$
To reconstruct $B_{i,k}$  from 
$w_1(V,\A,\theta,x)$ given for all $(\theta,x)\in {\cal V}_{i,k}$, we give the following scheme which is based on the formula (1.12c) 
(${\cal V}_{i,k}$ is defined in (1.11)). 
The formula (1.12c) gives the value of all  integrals of $B_{i,k}$ over any straight line of any two-dimensional
affine plane $Y$ whose tangent vector space is $Y_{i,k}=\{(x_1',\ldots,x_d')\in\R^d|x_j'=0,j\neq i,j\neq k\}$. Now, to reconstruct $B_{i,k}$ at a 
point $x'\in \R^d$ we consider in $\R^d$ a two-dimensional plane $Y$ containing $x'$ and whose tangent vector space is $Y_{i,k}$. We interpret
$T\S^{d-1}$ as the set of all rays in $\R^d$ and we consider in $T\S^{d-1}$ the subset $T\S^1(Y)$ which is the set of all rays lying in
$Y$. Then we restrict $PB_{i,k}$ on $T\S^1(Y)$ and reconstruct $B_{i,k}(x')$ from these data 
using methods of reconstruction of $f$ from $Pf$ for $d=2$. 
(We can also use the formula (1.12b) for reconstruction of  $B_{i,k}$ from $w_1(V,\A,\theta,x)$ given for all $(\theta,x)\in T\S^{d-1}$.)

\vskip 2mm

To prove Proposition 1.2 we will first prove Proposition 5.2 and Proposition 5.3 given below.

Let $\A\in C^2(\R^d,\R^d)$ satisfy (5.1). We define a function $\tilde{w_4}(\A):\R^d\b\{0\}\times\R^d\to \R^d$ by 
$\tilde{w_4}(\A)(y,x)=|y| w_4(\A,{y\over|y|}, x-{x.y\over |y|^2}y)$ for $x\in\R^d,\ y\in \R^d\b\{0\}.$ From (5.1) and (5.3b), it follows
that 
$$
\tilde{w_4}(\A)(y,x)=\int_{-\infty}^{-xy\over |y|^2}\!\int_{-\infty}^\tau\!\!\!\!\! B(x+\sigma y)yd\sigma d\tau
-\int^{+\infty}_{-xy\over |y|^2}\!\!\!\int^{+\infty}_\tau\!\!\!\!\! B(x+\sigma y)y d\sigma d\tau,\eqno(5.11)
$$ 
and $\tilde{w_4}(\A)=(\tilde{w_4}(\A)_1,..,\tilde{w_4}(\A)_d)\in C^1(\R^d\b\{0\}\times\R^d,\R^d).$
\vskip 2mm

\noindent{\bf Proposition  5.2.} 
{\it Let $\A \in C^2(\R^d,\R^d)$ satisfy} (5.1), {\it then  
$B$ defined by} (5.2) 

\noindent {\it satisfies:
$$
\eqalignno{
\sum_{j=1}^d\theta_j \left[\theta_kPB_{i,j}(\theta,x)
\right.-&\left.\theta_iPB_{k,j}(\theta,x)\right]
-PB_{i,k}(\theta,x)=&(5.12{\rm a})\cr
&{\pa\over\pa x_i}\tilde{w_4}(\A)_k(\theta,x)-{\pa\over\pa x_k}\tilde{w_4}(\A)_i(\theta,x),
}
$$
$$
\sum_{j=1}^d\theta_j \left[\theta_kPB_{i,j,l}(\theta,x)
-\theta_iPB_{k,j,l}(\theta,x)\right]
-PB_{i,k,l}(\theta,x)=\tilde{w_4}(\A)_{i,k,l}(\theta,x),
\eqno(5.12{\rm b})
$$
for $(\theta,x)\in T\S^{d-1},\ i,k,l=1..d,$ 
where $P$ denotes the X-ray transform  and where 
$\tilde{w_4}(\A)_{m,n,l}(\theta,x)=
{\pa\over\pa x_l}\left({\pa\over\pa x_m}\tilde{w_4}(\A)_n-{\pa\over\pa x_n}\tilde{w_4}(\A)_m\right)(\theta,x),$ 
$B_{m,n,l}(x)=$ 

\noindent ${\pa\over \pa x_l}B_{m,n}(x),$ for $\theta\in \S^{d-1},$ $x\in \R^d,$ $m,n=1..d$.
}
\vskip 2mm

\noindent{\bf Remark 5.1.} Let $d=3,$ $l=1..d.$ 
Formula (5.12b) gives in fact
$$
\eqalignno{
\theta\star(-PB_{2,3,l}(\theta,x),PB_{1,3,l}(\theta,x),
-PB_{1,2,l}(\theta,x))=\cr
\theta\star(\tilde{w_4}(\A)_{2,3,l}(\theta,x),-\tilde{w_4}(\A)_{1,3,l}(\theta,x),\tilde{w_4}(\A)_{1,2,l}(\theta,x)),&&(5.13)
}
$$
for any $(\theta,x)\in T\S^2$ where $\star$ denotes the usual scalar product on $\R^3.$
\vskip 2mm

\noindent{\it Proof of Proposition} 5.2.
Under the conditions (5.1), from (5.11) it follows that 
$$
\eqalignno{
{\pa\over\pa x_k}\tilde{w_4}(\A)_i(y,x)=&
-{y_k\over |y|^2}\sum_{j=1}^dy_j\int_{-\infty}^{+\infty}B_{i,j}(\sigma y+x)d\sigma d\tau&(5.14)\cr
&+\sum_{j=1}^dy_j\left\lbrace\int_{-\infty}^{-{x\star y\over |y|^2}}\int_{-\infty}^\tau 
{\pa\over \pa x_k}B_{i,j}(\sigma y+x)d\sigma d\tau\right.\cr
&\left.-\int^{+\infty}_{-{x\star y\over |y|^2}}\int^{+\infty}_\tau {\pa\over
\pa x_k}B_{i,j}(\sigma y+x)d\sigma d\tau\right\rbrace
}
$$
for any $(y,x)\in \R^d\b\{0\}\times\R^d$ and $i,k=1..d$.
Let $i,k,l=1..d$ be fixed.
From (5.14) it follows that
$$
\eqalignno{
&{\pa\over\pa x_k}\tilde{w_4}(\A)_i(\theta,x)-{\pa\over\pa x_i}\tilde{w_4}(\A)_k(\theta,x)\cr
=&-\theta_k\sum_{j=1}^d\theta_j\int_{-\infty}^{+\infty}B_{i,j}(\sigma\theta+x)d\sigma d\tau
+\theta_i\sum_{j=1}^d\theta_j\int_{-\infty}^{+\infty}B_{k,j}(\sigma\theta+x)d\sigma d\tau\cr
&+\sum_{j=1}^d\theta_j\left\lbrace\int_{-\infty}^{0}\int_{-\infty}^\tau [
{\pa\over \pa x_k}B_{i,j}(\sigma\theta+x)-{\pa\over \pa x_i}B_{k,j}(\sigma\theta+x)]d\sigma d\tau\right.\cr
&\left.-\int^{+\infty}_{0}\int^{+\infty}_\tau [
{\pa\over \pa x_k}B_{i,j}(\sigma\theta+x)-{\pa\over \pa x_i}B_{k,j}(\sigma\theta+x)]d\sigma d\tau\right\rbrace
&(5.15)
}
$$
for $x\in \R^d,\ \theta\in \S^{d-1},\ \theta=(\theta_1,\ldots,\theta_d).$
From (5.1) and (5.2) it follows that
$$
{\pa\over \pa x_k}B_{i,j}(x)-{\pa\over \pa x_i}B_{k,j}(x)={\pa\over \pa x_j}B_{i,k}(x),\ x\in\R^d,\ j=1..d.\eqno(5.16)
$$
Let $\theta\in \S^{d-1}$ be fixed. Using (5.15), (5.1) and (5.16) we obtain (5.12a). 
Under conditions (5.1), the function $h_{i,k,\theta}$ which is  defined by $ h_{i,k,\theta}(x)=
{\pa\over\pa x_k}\tilde{w_4}(\A)_i(\theta,x)$ $-{\pa\over\pa x_i}\tilde{w_4}(\A)_k(\theta,x),x\in\R^d,$ satisfies $h_{i,k,\theta}\in
C^1(\R^d,\R)$ and (5.12b) follows immediatly from (5.12a).

Proposition 5.2 is proved.
\hfill $\sqcap\hskip -2.32mm\sqcup$
\vskip 2mm

\noindent{\bf Proposition 5.3.} 
{\it Let $\A \in C^2(\R^d,\R^d)$ satisfy} (5.1){\it.} 

i. {\it if $d=2$ then $w_4(\A,\theta,x)$ given for all $(\theta,x)\in T\S^{d-1}$ does not determine uniquely the magnetic field $B$ 
defined by} (5.2),
 
ii. {\it if $d\ge3$ then $w_4(\A,\theta,x)$ given for all $(\theta,x)\in T\S^{d-1}$ uniquely determines the magnetic field $B$ 
defined by} (5.2). {\it In addition the following formulas are valid:
if $d=3$ then
$$
\eqalignno{
(-{\cal F}B_{2,3,l}(0),{\cal F}B_{1,3,l}(0),-{\cal F}B_{1,2,l}(0))
=&&(5.17{\rm a})\cr 
(2\pi)^{-3/2}\sum_{j=1}^3\left[\theta^j\star\left(\int_{\Pi_{\theta^j}}\tilde{w_4}(\A)_{2,3,l}(\theta^j,y)dy, 
-\int_{\Pi_{\theta^j}}\tilde{w_4}(\A)_{1,3,l}\right.\right.&(\theta^j,y)dy,\cr
\left.\left.\int_{\Pi_{\theta^j}}\tilde{w_4}(\A)_{1,2,l}(\theta^j,y)dy\right)\right]\theta^j,
}
$$
for any orthonormal basis $(\theta^1,\theta^2,\theta^3)$ and where $\Pi_{p'}$ is the vector plane $\{y\in \R^3|y\star p'=0\}$ for
$p'\in \R^3\b\{0\},$ and ${\cal F}$ denotes the classical Fourier transform on $L^1(\R^3);$ 
$$
\eqalignno{
(-{\cal F}B_{2,3,l}(p),{\cal F}B_{1,3,l}(p),-{\cal F}B_{1,2,l}(p))
=&&(5.17{\rm b})\cr 
(2\pi)^{-3/2}\sum_{j=1}^2\left[\theta^j_p\star\left(\int_{\Pi_{\theta^j_p}}e^{-iy\star p}\tilde{w_4}(\A)_{2,3,l}(\theta^j_p,y)dy,
\right.\right.&-\int_{\Pi_{\theta^j_p}}e^{-iy\star p}\tilde{w_4}(\A)_{1,3,l}(\theta^j_p,y)dy,\cr
\left.\left.\int_{\Pi_{\theta^j_p}}e^{-iy\star p}\tilde{w_4}(\A)_{1,2,l}(\theta^j_p,y)dy\right)\right]\theta^j_p,
}
$$
for $p\in \R^3\b\{0\}$ and any orthonormal family $\{\theta^1_p,\theta^2_p\}$ of the plane $\Pi_p$ (and where $i=\sqrt{-1}$);

if $d\ge 4$ then
$$
PB_{j,k}(\theta,x)={\pa\over\pa x_k}\tilde{w_4}(\A)_j(\theta,x)-{\pa\over\pa x_j}\tilde{w_4}(\A)_k(\theta,x)\eqno(5.17{\rm c})
$$
for $(\theta,x)\in \tilde{\cal V}_{j,k}$, where $\tilde{\cal V}_{j,k}$ is the $(2d-4)$-dimensional manifold $\{(\theta,x)\in
T\S^{d-1}|\theta$ $=(\theta_1,...,\theta_d), \theta_j=\theta_k=0\}.$
}
\vskip 2mm

\noindent{\it Proof of Proposition} 5.3.
We first prove the item (i). Let $\xi\in C^1(\R^+,\R)$ be such that
$$
\A(x)=(-x_2\xi(|x|^2),x_1\xi(|x|^2)),\ x\in \R^2,
$$
satisfies (5.1) and
$$
B(x)=\pmatrix{
0&2|x|^2\xi'(|x|^2)\cr
-2|x|^2\xi'(|x|^2)& 0\cr
}\not\equiv 0
$$
(e.g. $\xi(t)={1\over (1+t)^\sigma},t\in \R^+,\ \sigma>1$ or $\xi(t)=e^{-t},\ t\in \R^+$).
We define 
$
w_5(\A,\theta,x)=\int_{-\infty}^0\int_{-\infty}^\tau\! 2|\sigma\theta+x|^2\xi'(|\sigma\theta+x|^2)d\sigma d\tau
-\int^{+\infty}_0\int^{+\infty}_\tau\! 2|\sigma\theta+x|^2\xi'(|\sigma\theta+x|^2)d\sigma d\tau,\
$
for $(\theta,x)\in T\S^{d-1}.$ Let $(\theta,x)\in T\S^{d-1}$ be fixed. Using $|\sigma\theta+x|^2=\sigma^2+|x|^2$ we obtain $w_5(\A,\theta,x)=0$. 
From this equality and (5.3b) it follows that
$
w_4(\A,\theta,x)=w_5(\A,\theta,x)(\theta_2,-\theta_1)=0
$
($\theta=(\theta_1,\theta_2)$).
The item (i) is proved.

We prove the item (ii). We shall distinguish the case $d=3$ from the case $d\ge 4.$

First let $d\ge 4$.
Let $j,k=1..d$ be fixed, $j\neq k$. Formula (5.12a) implies (5.17c).  
Let $x'\in \R^d.$
As $d\ge 4$, the dimension of the vector space $H_{j,k}=\{(x_1,\ldots,x_d)$ $\in\R^d|x_j=x_k=0\}$ is greater than or equal to 2. Let $\{e_1,e_2\}$ be an
orthonormal family of $H_{j,k}.$ Let $Y$ be the affine plane of $\R^d$ which passes through $x'$ and whose tangent vector space is the vector space
spanned by $\{e_1,e_2\}.$ From (5.17c), it follows that the integral of $B_{j,k}$ over any straight line of $Y$ is known from 
${\pa\over\pa x_k}\tilde{w_4}(\A)_j(\theta,x)-{\pa\over\pa x_j}\tilde{w_4}(\A)_k(\theta,x)$ given for all $(\theta,x)\in
\tilde{\cal V}_{j,k}$. Thus using
results on inversion of the X-ray transform $P$ (see [GGG], [Na] and [No1]), we obtain that 
${B_{j,k}}_{|Y}$ can be reconstructed from ${\pa\over\pa x_k}\tilde{w_4}(\A)_j(\theta,x)-{\pa\over\pa x_j}\tilde{w_4}(\A)_k(\theta,x)$ given for all $(\theta,x)\in
\tilde{\cal V}_{j,k}$ 
(where ${B_{j,k}}_{|Y}$ denotes the restriction of $B_{j,k}$ to $Y$). Hence $B_{j,k}(x')$ can be reconstructed from 
${\pa\over\pa x_k}\tilde{w_4}(\A)_j(\theta,x)-{\pa\over\pa x_j}\tilde{w_4}(\A)_k(\theta,x)$ given for all $(\theta,x)\in
\tilde{\cal V}_{j,k}$.

Then let $d=3$ and let $l =1..3$ be fixed.
Under conditions (5.1),  $B_{j,k,l}\in L^1(\R^3),$ for $ j,k=1..3.$ Let $p\in\R^3$ be fixed. From (5.12b) and (5.13) we obtain
$$
\eqalignno{
\theta\star\left(-{\cal F}B_{2,3,l} (p),{\cal F}B_{1,3,l} (p),
-{\cal F}B_{1,2,l} (p)\right)=&&(5.18)\cr
(2\pi)^{-3/2}\theta\star\left(\int_{\Pi_{\theta}}e^{-iy\star p}\tilde{w_4}(\A)_{2,3,l}(\theta,y)dy,
\right.&-\int_{\Pi_{\theta}}e^{-iy\star p}\tilde{w_4}(\A)_{1,3,l}(\theta,y)dy,\cr
\left.\int_{\Pi_{\theta}}e^{-iy\star p}\tilde{w_4}(\A)_{1,2,l}(\theta,y)dy\right)
}
$$
for any $\theta\in \S^2,\ \theta\star p=0$.
The formula (5.18) implies (5.17a). 
To prove that (5.18) also implies (5.17b), we shall use the following 
\vskip 2mm
\noindent {\bf Lemma 5.1.} {\it  Under the conditions} (5.1), $(-{\cal F}B_{2,3,l} (p),{\cal F}B_{1,3,l} (p),
-{\cal F}B_{1,2,l} (p))\star p=0,$ {\it for} $p\in \R^3.$
\vskip 2mm
Lemma 5.1 and (5.18) imply (5.17b). 

Let $m,n=1,2,3$ $m\neq n.$ Using the injectivity of the Fourier transform and (5.17a) and (5.17b), we obtain that $B_{m,n,l}$ is uniquely determined by 
$w_4(\A,\theta,x)$ given for all $(\theta,x)\in T\S^{d-1}.$ Since $B_{m,n}$ vanishes at infinity, we deduce that $B_{m,n}$ is uniquely determined by 
$w_4(\A,\theta,x)$ given for all $(\theta,x)\in T\S^{d-1}.$

Proposition 5.3 is proved.
\hfill $\sqcap\hskip -2.32mm\sqcup$
\vskip 2mm
\noindent{\it Proof of Lemma} 5.1.
We define $\lambda:\R^3\to \R$ by 
$$
\lambda(p)=\left(-{\cal F}B_{2,3,l} (p),{\cal F}B_{1,3,l} (p),
-{\cal F}B_{1,2,l} (p)\right)\star p,\ p=(p_1,p_2,p_3)\in \R^3.\eqno(5.19)
$$ 

Now we shall use the tempered distributions and ${\cal F}$ shall denotes the Fourier transform of tempered distributions. 
Under conditions (5.1), $\A_i$ defines a tempered distribution of ${\cal S}'(\R^3).$

%We recall that for any tempered distribution $u$ on ${\cal S}(\R^3)$ we have the following properties
%$$
%{\cal F}D^{\sigma}u=x^{\sigma}u {\rm \ in\ } {\cal S}'(\R^3), 
%$$
%where $D^{\sigma}={1\over I}^{|\sigma|}{\pa^{\sigma_1}\over \pa x_1^{\sigma_1}}..{\pa^{\sigma_3}\over \pa x_3^{\sigma_3}},$
%$\sigma=(\sigma_1,..,\sigma_3)\in \N\union\{0\}^3,$ $x=(x_1,..,x_3)\in \R^3.$
From (5.2) and (5.19) it follows that
 
$$
\eqalignno{
<\lambda(p),\phi>=&<p_1p_lp_2{\cal F} \A_3-p_1p_lp_3{\cal F} \A_2-p_2p_lp_1{\cal F} \A_3\cr
&+p_2p_lp_3{\cal F} \A_1+
p_3p_lp_1{\cal F} \A_2-p_3p_lp_2{\cal F} \A_1,\phi>\cr
=&0&(5.20)
}
$$
for $\phi \in {\cal S}(\R^3).$
Since $B_{m,n,l}\in L^1(\R^3)$ , ${\cal F}B_{m,n,l}$ is a continuous function on $\R^3$ for any $m,n=1,2,3$. Thus $\lambda$ is continuous
on $\R^3$. From the continuity of $\lambda$ and (5.20), it follows that $\lambda\equiv 0.$ 

Lemma 5.1 is proved. 
\hfill $\sqcap\hskip -2.32mm\sqcup$
\vskip 2mm

Now we are ready to prove Proposition 1.2.
We note that
$$
w_4(\A,\theta,x)={1\over 2}(w_2(V,\A,\theta,x)-w_2(V,\A,-\theta,x)),\eqno(5.21{\rm a})
$$
$$
\eqalignno{
PV(\theta,x)\theta+&\int^{+\infty}_0\!\!\!\int^{+\infty}_\tau\!\!\!\nabla V(s\theta+x)dsd\tau
-\int_{-\infty}^0\!\int_{-\infty}^\tau\!\!\! \nabla V(s\theta+x)dsd\tau&(5.21{\rm b})\cr
&={1\over 2}(w_2(V,\A,\theta,x)+w_2(V,\A,-\theta,x)),\cr
}
$$
for $(\theta,x)\in T\S^{d-1}.$
The formulas (5.21), Proposition 1.1 of [Jo] and Proposition 5.3 imply Proposition 1.2.
\hfill $\sqcap\hskip -2.32mm\sqcup$
\vskip 2mm

\noindent{\bf Remark 5.2.} If we replace the conditions (5.1) and the formula (5.2) by the conditions (4.11b), (4.11c) and (4.11d), then 
Propositions 5.1, 5.2, 5.3 and Lemma 5.1 
still hold.
To prove Propositions 5.1, 5.2 and 5.3 under the conditions (4.11b), (4.11c) and (4.11d), we use the transversal gauge (given by (4.12)) 
and we follow Proof of Propositions 5.1, 5.2 and 5.3 under the conditions (5.1).
To prove Lemma 5.1 under the conditions (4.11b), (4.11c) and (4.11d), we use the transversal gauge $\A$ (given by (4.12)) and we note that
(4.13a) implies that
$\A_i$ defines a tempered distribution on ${\cal S}(\R^d)$ for $i=1..d,$ and we follow Proof of Lemma 5.1 under the conditions (5.1). 

\vskip 1cm

\noindent{\sect References}
\vskip 4mm

{\parindent=-1.2cm 
\leftskip=-\parindent 
\leavevmode \hbox to 1.2cm {[A]\hfill}S. Arians, Geometric approach to inverse scattering for the Schr\" odinger equation with magnetic and electric potentials, {\it J.
Math. Phys.} {\bf 38}(6), 2761-2773 (1997). 

\leavevmode \hbox to 1.2cm {[E]\hfill}A. Einstein, \" Uber das Relativit\" atsprinzip und die aus demselben gezogenen Folgerungen, 
{\it Jahrbuch der Radioaktivit\" at und Elektronik} {\bf 4},  411-462 (1907).

\leavevmode \hbox to 1.2cm {[ER1]\hfill}G. Eskin, J. Ralston, Inverse scattering problem for the Schr\" odinger equation with magnetic potential at a fixed energy, {\it Comm.
Math. Phys.} {\bf  173}, 199-224 (1995). 

\leavevmode \hbox to 1.2cm {[ER2]\hfill}G. Eskin, J. Ralston, Inverse scattering problems for the Schr\" odinger operators with magnetic and electric potentials, {\it IMA vol.
Math. Appl.} {\bf 90}, 147-166 (1997). 

\leavevmode \hbox to 1.2cm {[EW]\hfill}V. Enss, R. Weder, Inverse potential scattering: a geometrical approach, in {\it Mathematical Quantum Theory II: Schr\" odinger
operators} (Feldman, J., Froese, R. and Rosen, L., eds.), CRM Proc. Lecture Notes 8, Amer. Math. Soc., Providence, R.I., 1995. 

\leavevmode \hbox to 1.2cm {[F]\hfill}L.D. Faddeev, Uniqueness of solution of the inverse scattering problem, {\it Vestnik. Leningrad. Univ.} {\bf
11}(7), 126-130 (1956).

\leavevmode \hbox to 1.2cm {[GGG]\hfill}I.M. Gel'fand, S.G. Gindikin, M.I. Graev, Integral geometry
in affine and projective spaces, {\it Itogi Nauki i Tekhniki, Sovr. Prob. Mat.}
{\bf 16}, 53-226 (1980) (Russian).

\leavevmode \hbox to 1.2cm {[GN]\hfill}M.L. Gerver, N. S. Nadirashvili, Inverse problem of mechanics at high energies, 
{\it Comput. Seismology} {\bf 15}, 118-125 (1983) (Russian).

\leavevmode \hbox to 1.2cm {[Ha]\hfill}G. Hachem, The Faddeev formula in the inverse scattering for Dirac operators, {\it Helv. Phys. Acta} {\bf 72}(5\&6), 301-315 (1999).

\leavevmode \hbox to 1.2cm {[HN]\hfill}G.M. Henkin, R.G. Novikov, A multidimensional inverse problem in quantum and acoustic scattering, {\it Inverse
Problems} {\bf 4}, 103-121 (1988).  

\leavevmode \hbox to 1.2cm {[I]\hfill}H. T. Ito, High-energy behavior of the scattering amplitude for a Dirac operator, {\it Publ. RIMS, Kyoto Univ.}  {\bf 31}, 1107-1133
(1995).

\leavevmode \hbox to 1.2cm {[Jo]\hfill}A. Jollivet, On inverse scattering for the multidimensional relativistic Newton equation at 
high energies, {\it J. Math. Phys.} (2006), to appear (2005 

\noindent preprint, /math-ph/0502040).

\leavevmode \hbox to 1.2cm {[Ju]\hfill}W. Jung, Geometric approach to inverse scattering for Dirac equation, {\it J. Math. Phys.} {\bf 38}(1), 39-48 (1997).

\leavevmode \hbox to 1.2cm {[LL1]\hfill}L.D. Landau, E.M. Lifschitz, {\it Mechanics}, Pergamon Press Oxford, 1960.

\leavevmode \hbox to 1.2cm {[LL2]\hfill}L.D. Landau, E.M. Lifschitz, {\it The Classical Theory of Fields}, Pergamon Press New York, 
1971.

\leavevmode \hbox to 1.2cm {[Na]\hfill}F. Natterer, {\it The Mathematics of Computerized Tomography}, Teubner 

\leavevmode \hbox to 1.2cm {\hfill}Stuttgart and  Wiley Chichester, 1986.

\leavevmode \hbox to 1.2cm {[Ni]\hfill}F. Nicoleau, A stationary approach to inverse scattering for Schr\" odinger operators with 
first order perturbation, {\it Commun. partial
differ. equ.} {\bf 22} (3\&4), 527-553 (1997). 

\leavevmode \hbox to 1.2cm {[No1]\hfill}R.G. Novikov, Small angle scattering and X-ray transform
in classical mechanics, {\it Ark. Mat.} {\bf 37},  141-169 (1999).

\leavevmode \hbox to 1.2cm {[No2]\hfill}R.G. Novikov, The $\overline\partial$-approach to approximate inverse scattering at fixed energy 
in three dimensions, {\it IMRP Int. Math. Res. Pap.}  {\bf 6}, 287-349 (2005).

\leavevmode \hbox to 1.2cm {[R]\hfill}J. Radon, \" Uber die Bestimmung von Funktionen durch ihre Integralwerte l\" angs 
gewisser Mannigfaltigkeiten, {\it Ber. Verh. S\" achs. Akad. Wiss. Leipzig, Math.-Nat. K1} {\bf 69}, 262-267 (1917).

\leavevmode \hbox to 1.2cm {[S]\hfill}B. Simon, Wave operators for classical particle scattering, {\it Comm. Math. Phys.} {\bf 23}, 
37-48 (1971).

\leavevmode \hbox to 1.2cm {[Y]\hfill}K. Yajima, Classical scattering for relativistic particles, {\it J. Fac. Sci., Univ. Tokyo, 
Sect. I A} {\bf 29}, 599-611 (1982).

}

\vskip 8mm
\noindent A. Jollivet

\noindent Laboratoire de Math\'ematiques Jean Leray (UMR 6629)

\noindent Universit\'e de Nantes 

\noindent F-44322, Nantes cedex 03, BP 92208,  France

\noindent e-mail: jollivet@math.univ-nantes.fr

\end